\documentclass{emulateapj}
\usepackage{color}
\slugcomment{To be published by The Astrophysical Journal}

\shorttitle{Recurrent Buckling of Stellar Bars}
\shortauthors{Martinez-Valpuesta, Shlosman and Heller}

\begin{document}

\def\gtorder{\mathrel{\raise.3ex\hbox{$>$}\mkern-14mu
     \lower0.6ex\hbox{$\sim$}}}
\def\ltorder{\mathrel{\raise.3ex\hbox{$<$}\mkern-14mu
     \lower0.6ex\hbox{$\sim$}}}

\title{EVOLUTION OF STELLAR BARS IN LIVE AXISYMMETRIC HALOS: RECURRENT BUCKLING 
AND SECULAR GROWTH}

\author{Inma Martinez-Valpuesta$^{1,2}$} 

\author{Isaac Shlosman$^{2}$}
%\affil{Department of Physics, Astronomy \&  Mathematics\\
%        University of Hertfordshire, Hatfield, Herts, AL10 9AB, UK\\
%        email: {\tt martinez@pa.uky.edu}}
 
%\affil{Department of Physics \& Astronomy\\ University of Kentucky,
%        Lexington, KY 40506-0055, USA\\
%        email: {\tt shlosman@pa.uky.edu}}

%\and

\author{Clayton Heller$^{3}$}
%\author{Clayton Heller\altaffilmark{3}}
%\affil{Department of Physics\\ Georgia Southern University,
%        Statesboro, GA 30460, USA\\
%        email: {\tt cheller@georgiasouthern.edu}}

\altaffiltext{1}{Department of Physics, Astronomy \&  Mathematics,
        University of Hertfordshire, Hatfield, Herts, AL10 9AB, UK}
\altaffiltext{2}{Department of Physics \& Astronomy, University of Kentucky,
        Lexington, KY 40506-0055, USA}
\altaffiltext{3}{Department of Physics, Georgia Southern University,
        Statesboro, GA 30460, USA}

% July 16, 2004
% September 20, 2004
%\tableofcontents

\begin{abstract}
Evolution of stellar bars in disk galaxies is accompanied by dynamical
instabilities and secular changes. Following the vertical buckling 
instability, the bars are known to weaken dramatically and develop a pronounced 
boxy/peanut shape when observed edge-on. Using high-resolution $N$-body 
simulations of stellar disks embedded in live axisymmetric dark matter halos,
we have investigated the long-term changes in the bar morphology, specifically
the evolution of the bar size, its vertical structure and exchange of angular
momentum. We find that following the initial buckling, the bar resumes its
growth from deep inside the corotation radius and follows the Ultra-Harmonic
resonance thereafter. We also find that this secular bar growth triggers a
spectacular {\it secondary} vertical buckling instability which leads to the
appearance of characteristic boxy/peanut/X-shaped bulges. The secular bar
growth is crucial for the recurrent buckling to develop.
Furthermore, the secondary buckling is milder, persists over 
a substantial period of time, $\sim 3$~Gyr, and can have observational
counterparts. Overall, the stellar bars show recurrent behavior in their
properties and evolve by increasing their linear and vertical extents, both
dynamically and secularly. We also demonstrate explicitly that the prolonged
growth of the bar is mediated by continuous angular momentum transfer from
the disk to the surrounding halo, and that this angular momentum redistribution 
is resonant in nature --- a large number of lower resonances trap disk and
halo particles and this trapping is robust, in a broad agreement with the 
earlier results in the literature.
\end{abstract}

\keywords{galaxies: bulges --- galaxies: evolution --- galaxies: formation ---
galaxies: halos --- galaxies: kinematics and dynamics --- galaxies: spiral}
%%%%%%%%%%%%%%%%%%%%%%%%%%%%%%%%%%

%%%%%%%%%%%%%%%%%%%%%%%%%%%%%%%%%%%%%%%%%%%%%%%%%%%%%%%%%%%%%%%%%%
\section{Introduction}
%%%%%%%%%%%%%%%%%%%%%%%%%%%%%%%%%%%%%%%%%%%%%%%%%%%%%%%%%%%%%%%%%%

Observations and numerical modeling of galactic stellar bars have been
frequently accompanied with basic controversies about their origin and
evolution. While modern understanding of bar growth in a live and 
responsive environment is rooted in the angular momentum redistribution between 
the inner and outer disks, bulges and dark matter halos (Athanassoula 
2003), the efficiency of this process is hardly known and its details are still 
to be investigated. Recent efforts include but are not limited to the 
issues related to the bar lifetime cycles, gas-star interactions, bar
amplitudes and sizes, and bar slowdown (e.g., Bournaud \& Combes 2002; 
Valenzuela \& Klypin 2003; Shen \& Sellwood 2004; Weinberg 1985; Hernquist
\& Weinberg 1992; Debattista \& Sellwood 1998, 2000). The bigger issue of 
course is how the observational and theoretical aspects of bar
evolution fit within the emerging understanding of cosmological galaxy
evolution (e.g., Jogee et al. 2004; Elmegreen et al. 2004), specifically
the bar evolution in triaxial halos (El-Zant \& Shlosman 2002; Berentzen,
Shlosman \& Jogee 2005).

On one hand, early self-consistent models of numerical stellar bars 
have relied heavily on the Ostriker \& Peebles (1973) result which emphasized
the Maclauren sequence parameter $T/|W|$ --- the ratio of bulk
kinetic-to-gravitational energy as the threshold of bar instability. 
On the other hand, they revealed robustness of the bars --- once formed, the 
bars persisted (Athanassoula 1984 and refs. therein). The subsequent 
increase in the particle number above $N\sim 10^5$, switching from 2-D to 3-D 
models with responsive spheroidal components, and introduction of nonlinear 
physics tools in the orbital analysis have shown a much more complex bar
evolution and morphology than was anticipated originally (e.g., review by 
Athanassoula 2002a). This refers especially to the numerical confirmation
that live halos can indeed drive the bar instability rather than damp it
(Athanassoula \& Misiriotis 2002; Athanassoula 2003). Lastly, it is still
unclear to what extent and how 
closely the numerical bars correspond to their observed counterparts. This
issue exacerbates galaxy studies because overall, both theoretically and
observationally, 
the bars appear to be among the most important drivers of galactic evolution
across a wide range of spatial scales.

In this paper, we have revisited some aspects of a self-consistent 
evolution of stellar bars originating in {\it live} stellar disks embedded in
{\it live} dark matter halos by focusing on dynamical and secular 
changes\footnote{We refer to a dynamical evolution when changes develop
on the timescale of $\sim$ disk rotation, and to a secular evolution when they
develop on a much longer timescale of $\sim 10-100$ rotations.}
in these systems. The evolution of {\it numerical}
collisionless bars has been characterized so far in the literature by three
distinct phases --- the initial growth, the rapid vertical buckling 
and the prolonged quasi-steady regime, i.e., when the bars preserve their 
basic parameters (e.g., Sellwood \& Wilkinson 1993). However, some 
indication that bars can grow even in the last phase has been noticed already
in low-resolution 3-D models with live halos (e.g., Sellwood 1980). This
ability of
the bars to grow over extended period of time due to the momentum exchange
with the outer disk and especially with the halo has been confirmed recently
and analyzed in  
self-consistent 3-D simulations (Athanassoula 2003). Here we attempt to 
quantify this secular growth in terms of the bar size and its ellipticity, of
the angular momentum exchange, and of the ratio of vertical-to-radial
dispersion velocities. Moreover, we look into the corollaries of such
recurrent growth and find that it leads to additional and substantial 2-D and
3-D structural changes in the bar. We, therefore, discuss the observational 
consequences of this evolution.

Early in their growth stage, numerical stellar bars experience a 
dynamical instability --- the vertical buckling. The bars thicken profoundly, 
become more centrally-concentrated and acquire a characteristic peanut/boxy 
shape when seen edge-on (Combes et al. 1990; Pfenniger \& Friedli 1991; Raha
et 
al. 1991; Berentzen et al. 1998; Patsis, Skokos \& Athanassoula 2002a), while 
nearly dissolving the outer half of the bar, beyond the vertical inner Lindblad
resonance (Martinez-Valpuesta \& Shlosman 2004). This happens in live models
with both axisymmetric and mildly triaxial halos 
(Berentzen, Shlosman \& Jogee 2005). These boxy/peanut shapes are 
similar to bulge shapes observed in edge-on galaxies (e.g., Jarvis 1986;
Shaw 1987; Bureau \& Freeman 1999; Merrifield \& Kuijken 1999), which can be 
found in nearly half of all edge-on disk galaxies (L\"utticke, Dettmar \& 
Pohlen 2000). Although observed in numerical simulations a long time ago
(Combes \& Sanders 1981), the origin of boxy/peanut bulge shapes still
has two alternative explanations --- the well-known firehose instability 
(e.g., Toomre 1966; Raha et al. 1991; Merritt \& Sellwood 1994) and the
resonance heating (e.g., Combes et al. 1990; Pfenniger \& Friedli 1991; Patsis
et al. 2002b). These two views can be reconciled if buckling is responsible
for shortening the secular timescale of particle diffusion out of the disk 
plane and for accelerating the buildup of boxy/peanut bulges which proceeds 
on a much shorter dynamical timescale instead (Martinez-Valpuesta \& Shlosman 
2004).

However, is the buckling really necessary for a buildup of these boxy/peanut shaped
bulges? After all, even {\it imposing} vertical symmetry did not eliminate
this effect, albeit the buildup proceeded on a much longer timescale (Friedli
\& Pfenniger 1990). Where are the observational counterparts of these
asymmetric buckled bars? Due to a particular importance of the buckling
instability for the evolution of numerical bars and its plausible connection
to the buildup of the pronounced 3-D structure there, we have analyzed the bar
behavior during and following this instability. Specifically, we find that the
bars in a live environment are capable of recurrent growth, that the buckling 
instability is a recurrent event and that the buildup of the 3-D shape is not 
necessarily a dynamic phenomenon.

Much of the analysis of bar evolution and the accompaning
instabilities is implemented here by means of nonlinear orbit analysis, because
for such strong departures from axial symmetry the epicyclic approximation
cannot be relied upon --- being wrong quantitatively it frequently leads to
qualitative errors. We shall try to avoid the specific jargon associated with
this technique where possible. The angular momentum redistribution in the
model is quantified using orbital spectral analysis.

In section~2 we provide the details of numerical modeling and 
analysis. Section~3 describes the overall results of secular bar 
evolution. The 3-D bar orbital structure and the inferred vertical structure in
the bar are analyzed in section~4, and the resonant interaction between the
disk and the halo in section~5. Discussion and conclusions are given in
sections 6 and 7.

%%%%%%%%%%%%%%%%%%%%%%%%%%%%%%%%%%%%%%%%%%%%%%%%%%%%%%%%%%%%%%%%%%%%%%%%
\section{Numerical Tools and Modeling}
%%%%%%%%%%%%%%%%%%%%%%%%%%%%%%%%%%%%%%%%%%%%%%%%%%%%%%%%%%%%%%%%%%%%%%%

%%%%%%%%%%%%%%%%%%%%%%%%%%%%%%%%%%%%%%%%%%%%%%%%%%%%%%%%
\subsection{$N$-Body Simulations}

To simulate the stellar disk embedded in a live dark matter halo, we have
used version FTM-4.4 of the $N$-body code (Heller \& Shlosman 1994; 
Heller 1995) with $N = 10^5 - 1.1\times 10^6$. The gravitational forces 
are computed using Dehnen's (2002) {\tt falcON} force solver, a tree code 
with mutual cell-cell interactions and complexity {{\cal O}(N)}. It 
conserves momentum exactly and is about 10 times faster than an optimally 
coded Barnes \& Hut (1986) tree code. 

The initial density distribution is derived from the Fall \& Efstathiou 
(1980) disk-halo analytical model. The system is not in exact virial 
equilibrium, and therefore must be relaxed iteratively. The halo-to-disk 
mass ratio within 10~kpc is fixed to unity. The halo has a flat density 
core of 2~kpc to avoid excessive stochastic behavior associated with the 
central cusps (El-Zant \& Shlosman 2002). The disk is exponential and its 
radial and vertical 
scalelengths are taken as 2.85~kpc and 0.5~kpc, respectively. The disk 
and halo cut off radii are 25~kpc and 30~kpc and the initial circular 
velocity curves for disk and halo components and their sum are given in 
Fig.~1. The gravitational softening 
used is 160~pc and Toomre's parameter $Q = 1.5$. The adopted units are those
of $G=1$, mass $M=10^{11}~{\rm M_\odot}$ and distance $r=10$~kpc. 
This leads to the time unit of $\tau_{dyn}=4.7\times 10^7$~yrs and a velocity 
unit of 208~${\rm km~s^{-1}}$. The energy and 
angular momentum in the system are conserved to within approximately 1\% 
and 0.05\% accuracy, respectively. Above $N \sim 10^5$ our results appear 
to be reasonably independent of $N$. The model evolution presented here has 
$N=1.1\times 10^6$, with $8\times 10^5$ particles in the disk. 
 
\begin{figure}[ht!!!!!!]
\vbox to1.9in{\rule{0pt}{1.9in}}
\includegraphics{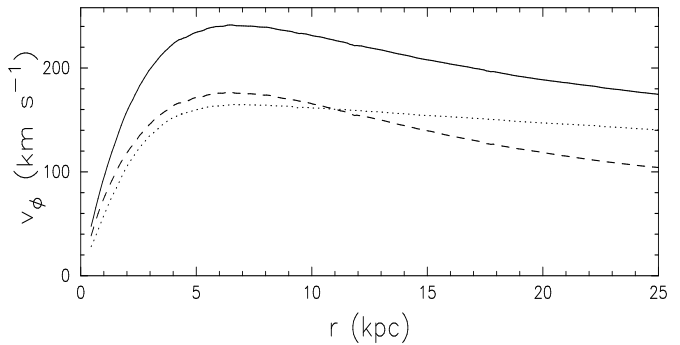}
\caption{Initial circular rotation velocities for the disk
(dashed lane) and halo (dotted lane) components. The total is given by the solid
lane. 
\label{fig:a2ampl}
}
\end{figure}

%%%%%%%%%%%%%%%%%%%%%%%%%%%%%%%%%%%%%%%%%%%%%%%%%%%%%%%%%%%%%%%%%%%%%%%%%%
\subsection{Orbital and Spectral Analysis}

The self-consistent evolution of stellar bars can be only understood by
studying their 3-D structure (e.g., Pfenniger \& Friedli 1991; Skokos, Patsis
\& Athanassoula 2002a,b). For such in-depth investigation we use the updated
algorithm described in Heller \& Shlosman (1996), which is based 
on a comprehensive search for periodic orbits in arbitrary gravitational
potentials, and display them in characteristic diagrams (section~4). 
These orbits close in the bar frame and provide the backbone for any meaningful
analysis of the bar properties. Periodic orbits characterize the overall
orbital structure of the bar phase space, because each of them traps a region
of phase space around it. These trapped orbits have shapes similar to the
shapes of the parent periodic orbit. The algorithm is run using the package
Parallel Virtual Machine (PVM), which distributes the search for
orbits among different processors and computes the orbits using adaptive step
size, with a relative accuracy of $10^{-7}$. We also calculate the stability
of these periodic orbits to estimate which orbits can be populated. We track
both the 2-D orbital families and the 3-D families bifurcating from planar
(i.e., equatorial) orbits at the (vertically) unstable gaps. The shapes of
these 3-D orbital families will contribute to the shape of the simulated bar,
when populated, including the evolving shape of the growing boxy/peanut bulge.

To quantify the disk--halo interaction and the angular momentum redsitribution
in the system, we have developed a package based on the orbital analysis
algorithm and which uses the Fast Fourier Transform (FFT) to find the main
orbital
frequencies --- the angular frequency $\Omega$, and the radial and vertical
epicyclic frequencies $\kappa$ and $\nu$ in the disk and the halo. The particle
distribution with the frequency ratio is then determined to find the
population of resonant orbits. Lastly the change in the angular momentum is
computed for each of these particles (see section~5).
 
%%%%%%%%%%%%%%%%%%%%%%%%%%%%%%%%%%%%%%%%%%%%%%%%%%%%%%%%%%%%%%%%%%%%%%%
\section{Results}
%%%%%%%%%%%%%%%%%%%%%%%%%%%%%%%%%%%%%%%%%%%%%%%%%%%%%%%%%%%%%%%%%%%%%%%%

We first describe the overall bar evolution during the simulation period
of $\tau \sim 14$~Gyr and analyze some of its more important behaviors. The 
bar develops in otherwise axisymmetric model during the first two-three 
disk rotations. Initially it has an axial ratio (flatness) $c/a \sim 0.1$, the 
same thickness as the disk, but this ratio increases with time by a
factor of $\sim 2$. The bar starts to brake against the outer disk and 
the halo as seen in Fig.~2b, reaching maximum strength at $\tau\sim 
1.4$~Gyr. At around $\tau\sim 1.8$~Gyr the bar experiences a vertical 
buckling instability which affects its 2-D and 3-D appearance. The outer 
part of the bar nearly dissolves, while overall the bar is weakened 
dramatically, as shown by the $m=2$ amplitudes, $A_2$, in Fig.~2a. 
Immediately following this buckling, the bar resumes its growth which 
saturates again at $\tau\sim 6$~Gyr. At this time 
$A_2$ in the outer part weakens again. The growth is resumed after 
$\tau\sim 7.5-8$~Gyr. A close inspection of the edge-on bar frames during time 
intervals of $1.8-2.8$~Gyr and $6-7.5$~Gyr and the analysis described in the 
next sections, reveal sufficient similarities between these two events --- 
both represent the vertical buckling instability in the bar, i.e., the breaking
of symmetry of the bar (Fig.~3). Such a recurrent buckling of bars has never
been reported in the literature. 

\begin{figure}[ht!!!!!!]
\vbox to3.3in{\rule{0pt}{3.3in}}
\includegraphics{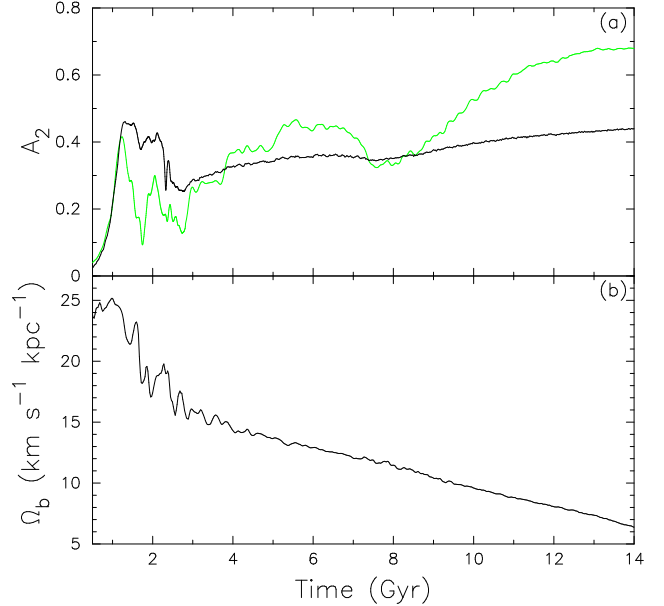}
\caption{{\it (a).} Evolution of bar $m=2$ amplitude, $A_2$, for 
$r=0-11$~kpc (thick solid) and $r=7-11$~kpc (thin solid) --- inner and outer
bar parts; {\it (b).} Bar pattern speed $\Omega_{\rm b}$. 
\label{fig:a2ampl}
}
\end{figure}

%%%%%%%%%%%%%%%%%%%%%%%%%%%%%%%%%%%%%%%%%%%%%%%%%%%%%%%%%%%%%%%%%%%%%%%%%%
\subsection{Secular Growth of Stellar Bars}

The stellar bar evolution presented here is characterized by substantial 
changes in the bar size and strength, and by changes in its 3-D shape (Fig.~3
and Animation Sequence~1). 
Determination of the live bar size in numerical simulations is not trivial. 
For example, Athanassoula \& Misiriotis (2002) and O'Neil \& Dubinski (2003) 
used a variety of methods and found that some of them give erroneous and 
unreliable results. The position of 
$A_2(r)$ maximum does not provide any meaningful estimate for the bar size 
because the contribution of higher harmonics, such as $m=4$ and 8, is
neglected. To quantify the bar size changes we have used two alternative
methods --- an ellipse fitting to the isodensity curves and the
characteristic orbital diagrams. The former method has been previously used to
detect and to characterize `observational' bars (e.g., Knapen et al. 2000;
Laine
et al. 2002; Hunt \& Malkan 2004). Its main deficiency when applied to 
numerical bars is an excessive noise for low-to-moderate $N$ and the resulting
distribution of the bar ellipticities, $\epsilon(r)$, being flat for young
unbuckled bars (Martinez-Valpuesta \& Shlosman 2004). However, at later times,
with the growth of the central mass concentration, this method becomes more
reliable. We find that a consistently reliable estimate of the bar size,
$r_{\rm bar}$, is the radius where $\epsilon(r)$ declines $\sim 15\%$ from its 
maximal value. The alternative and a {\it new} method used by us here relates 
the bar extent to 
the size of the maximal stable orbit of the main orbital family supporting the
bar (more in Section~4). Both methods produce consistent results with each
other.
 
\begin{figure*}[ht!!!!!!!!!!]
\vbox to6.8in{\rule{0pt}{6.1in}}
\includegraphics{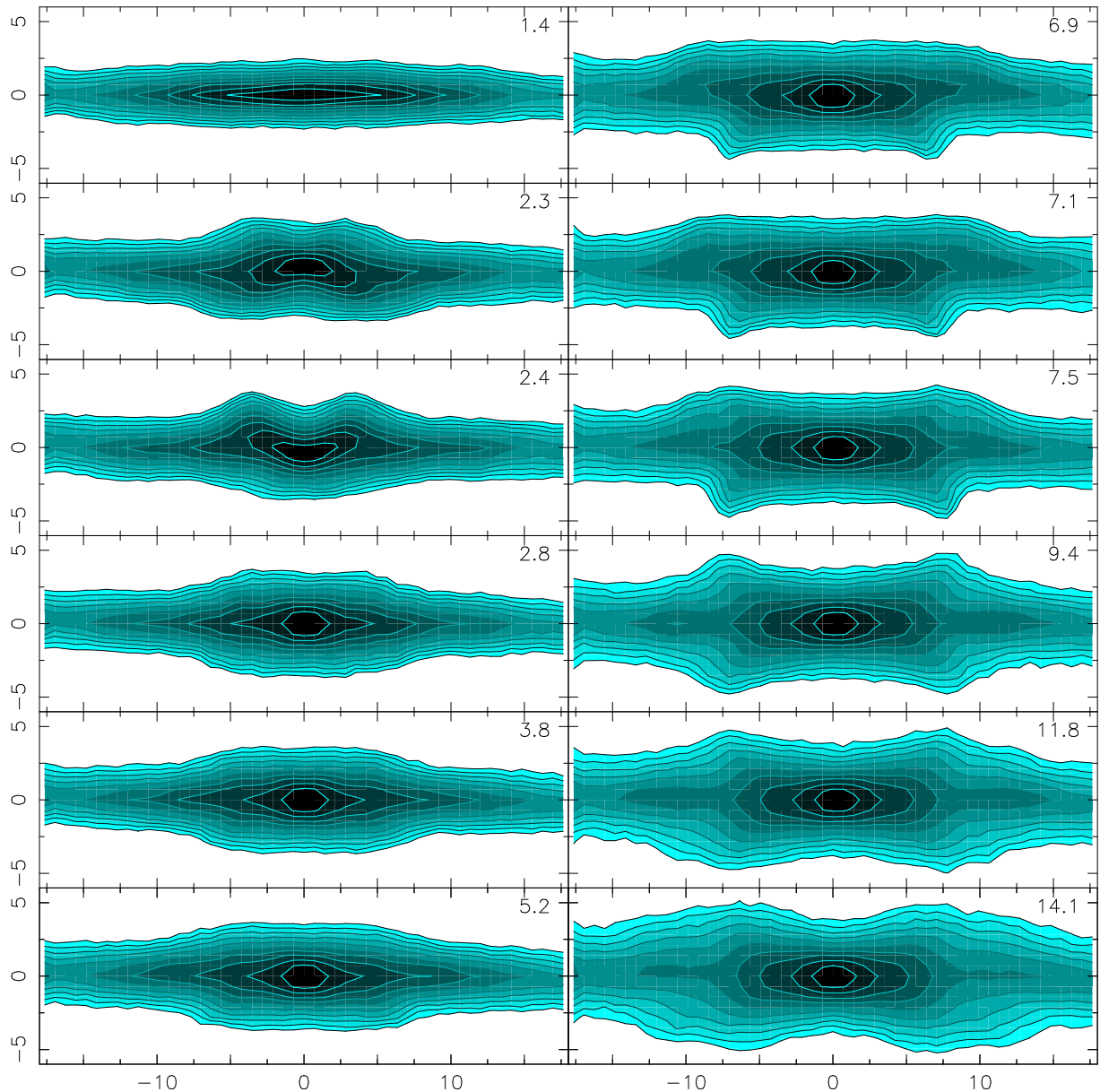}
\caption{Evolution of the vertical structure in the bar: edge-on view 
along the bar minor axis (see also Animation Sequence~1). The length is given 
in kpc and the values of the projected isodensity contours are kept unchanged
in all frames. The time in Gyrs is
given in the upper right corners. The maximal vertical asymmetries correspond 
to two recurrent bucklings, at $\tau\sim 2.4$~Gyr and at $\sim 7$~Gyr. Note, 
the bar flip-flop between $\tau=2.3$~Gyr and 2.4~Gyr; the persistent vertical 
asymmetry in the bar at $\tau=5.2-7.5$~Gyr; and the development of narrow 
features in the bar midplane outside its core of $\sim 8$~kpc, after $\tau\sim
9.4$~Gyr. Those correspond to {\it ansae} (`handles') in the face-on bar 
(see Fig.~11) and are observed in early type disk galaxies.
\label{fig:a2ampl}
}
\end{figure*}

The bar size evolution is shown in Fig.~4a, for both methods. It 
grows initially to 11~kpc, then buckles and shortens to 6~kpc. It then 
grows again to 13~kpc, where the growth stagnates due to the secondary 
buckling, for about 3~Gyr. After this the bar resumes its growth to about 
16~kpc. The size evolution in the $xy$ plane is accompanied by the 
vertical thickening of the bar. It does not stop after the first buckling 
instability, but continues gradually due to the vertical resonance scattering,
amplified by the recurrent buckling instability of the bar. The
accompanied boxy/peanut shaped bulge also grows with time (Fig.~3). 

\begin{figure}[ht!!!!!!]
\vbox to4.6in{\rule{0pt}{4.6in}}
\includegraphics{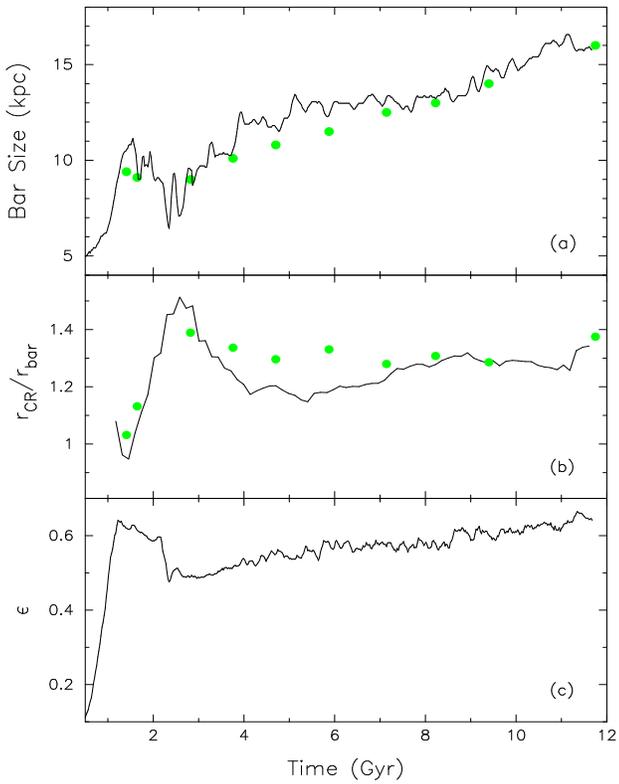}

\caption{{\it (a)} Evolution of the bar size (semimajor axis). The 
solid line represents the bar size obtained from the ellipse fitting to the 
isodensities. The filled dots correspond to the bar sizes obtained from the 
semimajor axis of the last stable $x_1$ periodic orbit which supports the bar,
obtained from Fig.~7 --- a new method introduced here (see Section~4 for 
details); {\it (b)} of bar
corotation-to-size ratio, and of {\it (c)} the maximal ellipticity of the bar,
$\epsilon = 1-b/a$ from the ellipse fitting. 
\label{fig:a2ampl}
}
\end{figure}

With the bar length, $r_{\rm bar}$, and its corotation radius, $r_{\rm CR}$, we
can quantify some of the dynamical characteristics of an evolving bar. The
ratio $r_{\rm CR}/r_{\rm bar}$ is shown in Fig.~4b. This ratio 
determines the shape of the offset dust lanes in barred galaxies which
delineate shocks in the gas flow (Athanassoula 1992). The observed shapes
constrain the ratio $r_{\rm CR}/r_{\rm bar}$ to 
$1.2\pm 0.2$. The modeled ratio typically falls within 
the required limits except during the first buckling when it is higher, 
$\sim 1.5$, as noticed already by Martinez-Valpuesta \& Shlosman (2004).
We shall discuss this issue in more detail in Section~6. 
 
%\newpage
%%%%%%%%%%%%%%%%%%%%%%%%%%%%%%%%%%%%%%%%%%%%%%%%%%%%%%%%%%%%%%%%%%%%%%%%%%
%%%%%%%%%%%%%%%%%%%%%%%%%%%%%%%%%%%%%%%%%%%%%%%%%%%%%%%%%%%%%%%%%%%%%%%%%%
\subsection{Recurrent Buckling of Stellar Bars}

The buckling is a 3-D phenomena and is most visible in the vertical $xz$-plane
(Fig.~3). To quantify the bar asymmetry, we 
have calculated the vertical $m=1$ mode amplitude $A_{1z}$ in the $xz$ plane
and follow the maximal 
distortion of the bar during its buckling periods. Two maxima
are apparent at $\sim 2.4$~Gyr and $\sim 7$~Gyr in Fig.~5. Note, that the
vertical asymmetry given by this Figure is building up slower and is weaker for 
the second buckling, i.e., from $\tau\sim 5$~Gyr to
$\tau\sim 7$~Gyr the amplitude is growing to $A_{1z} \sim 0.03$, compared to
$A_{1z} \sim 0.08$ at $\tau\sim 2.4$~Gyr. Fig.~3 shows this evolution in a more
graphical way --- while the first buckling affects mostly the central few kpc,
especially the bar's midplane, the second buckling is most prominent in the
outer bar range of 5--10~kpc and affects the midplane much less visibly. 

\begin{figure}[ht!!!!!!]
\vbox to3.3in{\rule{0pt}{3.3in}}
\includegraphics{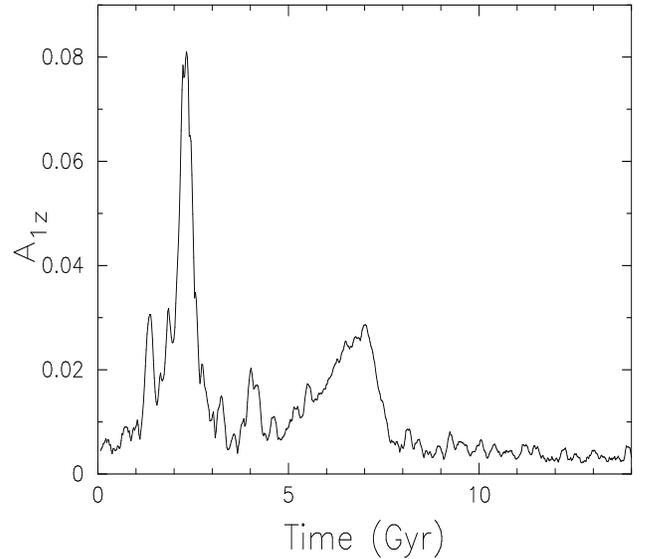}
\caption{Evolution of the vertical buckling amplitude in the bar,
$A_{1z}$, i.e. vertical $m=1$ mode in the $xz$ plane, integrated over 
$r\sim 0-12.5$~kpc, $-2$~kpc$<y<2$~kpc and $-\infty<z<\infty$ intervals.
\label{fig:a2ampl}
}
\end{figure}
 
The recurrent buckling can be detected in a number of ways, e.g., from $A_2$ in 
the $xy$-plane (Fig.~2a), as mentioned in section~3.1 and above, and from 
$A_{1z}$ in the $xz$ plane, which 
quantifies the breaking of vertical symmetry in the bar (Fig.~5). Each of the
coefficients emphasizes a different property of this instability. We are basically 
looking 
at the same phenomena at different times --- the first buckling extends over 
$\sim 1$~Gyr and the second one over $\sim 3$~Gyr. The changes in the 
orbital structure of the bar during the bucklings will be analyzed in
Section~4.  

We have commented in Section~1 that the buckling instability is a
collective breaking of a vertical symmetry in the bar. Toomre (1966) has shown
that the coupling between the vertical and radial degrees of motion is the
prime
driver of this instability and the main outcome of it is the equalizing of 
the velocity dispersion in the $xy$ plane with the vertical velocity
dispersion.\footnote{Normally, the kinetic energy of oscillations
about the equatorial plane, i.e. along the $z$-axis, is an adiabatic
invariant} 
This evolution is characterized by changes in vertical-to-radial velocity 
dispersion ratio $\sigma_{\rm z}^2/\sigma_{\rm r}^2$. Toomre estimated
the critical value for this ratio to lie at $\sim 0.1$ for a non-rotating 
plane-parallel slab,
Raha et al. (1991) at 0.06-0.3 for a 3-D stellar disk, and Sotnikova \& 
Rodionov (2005) at $\sim 0.6$ for the central regions embedded in
the hot halos. Sellwood (1996) have shown that Toomre's limit is violated for
many of his stellar models, some remaining  unstable up to 0.4. We plot this
ratio for two areas of the bar at two different times. For the first 
buckling (Fig.~6), the velocity dispersions are 
calculated in the central kpc, where the maximum effect is expected (e.g., 
Fig.~3). During this buckling, we observe first an increase in 
$\sigma_{\rm r}$ (the initial growth of the bar), and after 
$\sim 1$~Gyr a decrease in $\sigma_{\rm r}$ with a corresponding
increase in $\sigma_{\rm z}$. When $\sigma_{\rm z}^2/\sigma_{\rm r}^2$ drops
below 
$\sim 0.4$ the bar buckles and weakens. The buckling ends when 
$\sigma_{\rm z}^2/\sigma_{\rm r}^2$ increases to unity. The bar thickens and
grows and the vertical ILR moves gradually out. We expect and observe the 
maximal vertical asymmetry in the outer part of the bar during the second 
buckling, and therefore
calculate $\sigma_{\rm z}^2/\sigma_{\rm r}^2$ at around 7~kpc. Again, the 
gradual increase in $\sigma_{\rm z}$ and decrease in $\sigma_{\rm r}$ drive the 
ratio up from 0.4 to about unity, similar to the first buckling. For 
both bucklings we observe a very similar evolution in terms of the velocity 
dispersions and their ratios.

\begin{figure*}[ht!!!!!!]
\vbox to3.4in{\rule{0pt}{3.3in}}\includegraphics{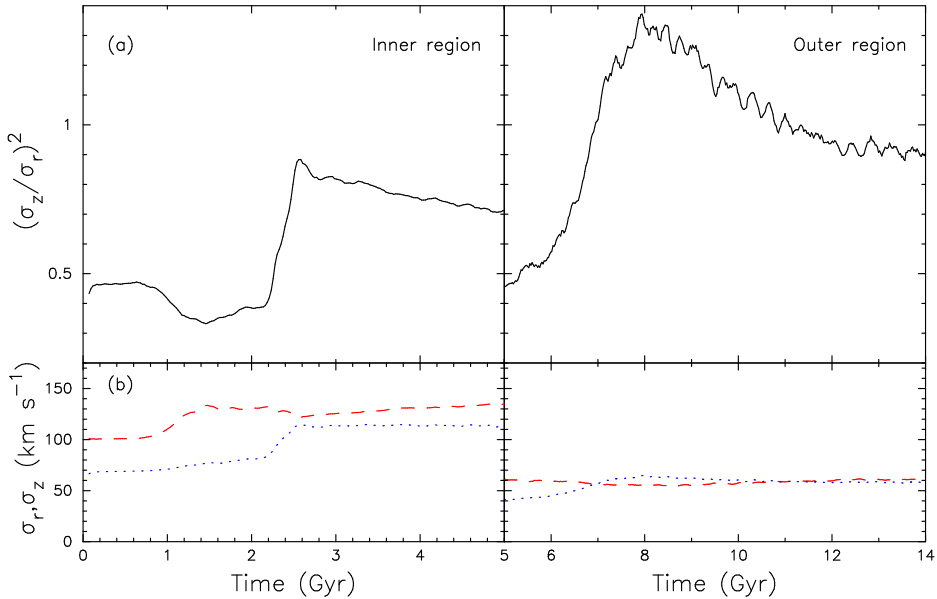}
\caption{Evolution of {\it (a)} the vertical-to-radial dispersion
velocity
ratio in the disk, $\sigma_{\rm z}^2/\sigma_{\rm r}^2$, within the central kpc
(left) and within a cylindrical shell of radius $7\pm 1$~kpc (right), and 
{\it (b)} separately, the radial, $\sigma_{\rm r}$ (dashed), and the vertical, 
$\sigma_{\rm z}$  (dotted), dispersion velocities within the same
region as in $(a)$. 
\label{fig:a2ampl}
}
\end{figure*}

So far we have shown that the bar buckles twice during the simulations: first 
time --- abruptly, fast and in the central part, and the second time ---
slower, less pronounced 
and in a different part of the bar. In both bucklings we observe the loss 
of symmetry in the vertical plane, the drop in $A_2$ and the equilizing of the 
vertical and radial velocity dispersions. After each buckling the bar 
becomes more symmetric, with the difference that after the first buckling 
the asymmetry starts to build anew, and after the second buckling ---  the
asymmetry is completely washed out.

To summarize, three main factors characterize  
the end of the buckling instability: $A_2$ starts to decrease, 
the asymmetry in the $rz$-plane decreases (given by $A_{1z}$), and  
$\sigma_{\rm z}^2/\sigma_{\rm r}^2\rightarrow 1$. Applying 
these conditions to our model, the end of the first 
buckling appears at $\sim 2.4-2.8$~Gyr and the end of the second buckling 
at $\sim 8$~Gyr.

Since the buckling leads to a sudden vertical thickening of the bar and 
is therefore characterized by particle injection above the disk 
plane, we can monitor this instability by following the particle 
distribution. When viewed along the bar minor axis, 
during the first buckling the bar bends and
develops a boxy/peanut shaped bulge, while during the second buckling the bar
acquires an asymmetric shape which leads to the appearance of an 
X-shaped bulge (Fig.~3). These shapes have a direct relationship to the
population of orbits trapped by the main family of periodic orbits in the bar
(Pfenniger \& Friedli 1991; Section~4 below). 

%%%%%%%%%%%%%%%%%%%%%%%%%%%%%%%%%%%%%%%%%%%%%%%%%%%%%%%%%%%%%%%%%%%%%%%%%%
\section{Analysis: Bar Orbital Structure Evolution}
%%%%%%%%%%%%%%%%%%%%%%%%%%%%%%%%%%%%%%%%%%%%%%%%%%%%%%%%%%%%%%%%%%%%%%%%%%%%

Stellar bars are 3-D objects which exhibit dynamical and secular evolutionary
trends both in their morphology and internal structure. In this section, we
study the evolution of the orbital structure in the bar by searching for the
{\it main} 2-D and 3-D families of orbits at various snapshots, with a
particular emphasis on the recurrent buckling periods. A comprehensive search
for the orbits is beyond the scope of this work.  We analyze the main
parameters of detected orbits and calculate their stability. This section uses
a specific terminology developed for nonlinear  orbital dynamics (e.g.,
Binney \& Tremaine 1987; Sellwood \& Wilkinson 1993). Reader unfamiliar with
this terminology can skip it and go directly to Section~5. We shall discuss
the results of the orbital analysis in the context of the bar evolution in
Section~6.  

The initial vertical axial ratio of the bar which develops in our models is
$c/a\sim 0.1$ (nearly a 2-D object), and grows dynamically and secularly to
$\sim 0.2$ over the Hubble time. We first analyze the bar's orbital structure
in
the equatorial plane, then add the 3-D effects accounting for the finite
thickness and shape of the bar. In doing so, we limit our analysis to the
periodic
orbits within $r_{\rm CR}$, which are largely responsible for the bar shape,
and to specific times of  bar evolution: $\tau=1.4$~Gyr (just prior to first
buckling), 2.8~Gyr (after the first buckling), 7.1~Gyr (during the second
buckling) and 11.8~Gyr (after the second buckling). The orbits are searched in
the potential symmetrized horizontally with respect to the four quadrants.
Fig.~5 confirms that the vertical symmetry is violated during the buckling
periods. The first buckling is fast and there is little meaning, therefore, to
calculate the orbits without some frame averaging --- the stars see only a
time-averaged potential. Hence, we resort to a vertically-symmetrised
potential at 2.8~Gyr. On the other hand, the second buckling is much more
gradual and we calculate the orbits in the actual `raw' potential at
$\tau=7.1$~Gyr to capture the persisting asymmetry. The third snapshot with
the bar being symmetric again is treated similarly to the first snapshot for
simplicity.

The extent of the orbital families and their stability are displayed by means
of characteristic diagrams. The $y$, $z$, or $\dot{z}$ intercept values with 
the $x=0$ plane are plotted with respect to their Jacobi integral, $E_{\rm J}$
(e.g., Binney \& Tremaine 1987). The Jacobi (energy) integral of motion is
conserved along any given orbit in the rotating bar frame. The orbits form
curves or families in the characteristic diagrams.  The actual trajectories
will not coincide exactly with these periodic orbits but may be `trapped' in
their vicinity. The properties of periodic orbits and their temporal changes,
therefore, reflect the bar structural evolution.  

% orbits2.8-11.8.ps
\begin{figure*}[ht!!!!!!!!!!!!!!!!!!!!!!!!!!!!!!!!!!!!!!!!!!]
\vbox to4.2in{\rule{0pt}{3.8in}}
\includegraphics{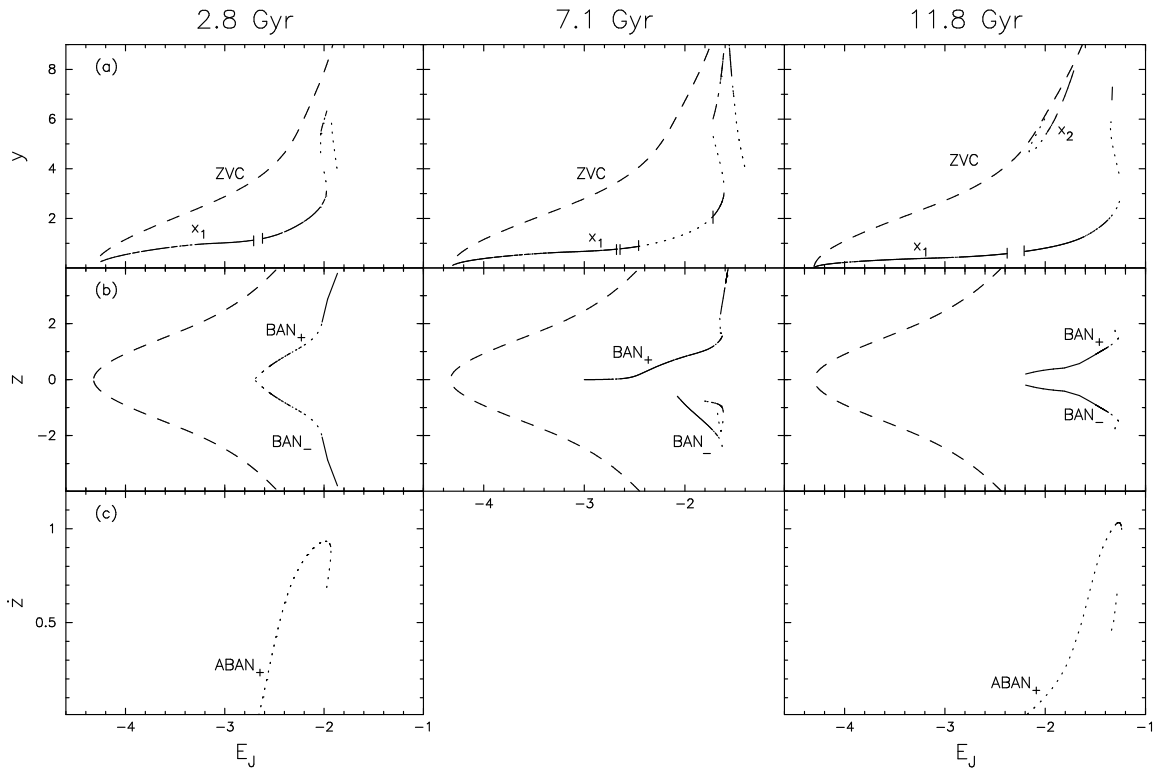}
\caption{Three snapshots of the bar orbital structure evolution. 
Left --- immediately after the first buckling, middle --- during the second 
buckling, right --- well after the second buckling. The $\tau=7.1$~Gyr
frame is non-symmetrized (vertically), while others are vertically symmetrized. 
{\it (a).} Characteristic diagrams in the $xy$-plane showing $y$-intersections
of regular orbits with the $x=0$ axis as a function of Jacobi energy of 
the orbits. The bar is oriented along the $x$-axis. Solid lines represent 
stable orbits, dotted --- unstable ones. The dashed lines show the ZVCs --- 
the zero velocity curves.
{\it (b).} Characteristic diagram in the $xz$-plane showing $z$-intersections
of regular orbits. Since 7.1~Gyr snapshot is non-symmetrized and  
the bar buckles in this plane, the distribution of the BAN 
orbits is also asymmetric, and most of the stable BAN orbits are ``$\frown$"
(see text). They have a larger curvature and extend further into the $z<0$
region leading to the bar asymmetry in the edge-on view (Fig.~3).
{\it (c).} Characteristic diagrams in the $xz$-plane showing the $\dot z$ 
families. The 7.1~Gyr snapshot is omitted because all the ABAN orbits are
unstable at this time and, therefore, difficult to calculate in unsymmetrized
potential.
\label{fig:orb2.8-11.8}
}
\end{figure*}

The most important single-periodic orbits in the bar midplane are those aligned
parallel or normal to the bar major axis, the so-called $x_1$ and
$x_2$ orbits (e.g., Contopoulos \& Papayannopoulos 1980; Binney \& Tremaine
1987). The $x_1$ constitutes the main family of orbits which support the bar
figure. While this family exists always within $r_{\rm CR}$, the $x_2$
appear only if the planar inner Lindblad resonance(s) (ILRs) are present.
Typically, numerical bars form with a pattern speed sufficiently high to avoid
the ILRs, at least in the beginning, and hence the $x_1$ family completely
dominates the bar midplane, short of the corotation region. 

The vertical shape of the bar is determined by the projection of the populated
3-D orbits onto the corresponding planes. We search for the
vertically-unstable gaps in the $x_1$ family. It is at these gaps where the 3-D
families bifurcate through $z$ and $\dot z$ bifurcations (e.g., Pfenniger
1984; Pfenniger \& Friedli 1991; Skokos et al. 2002a,b;
Patsis et al. 2002b). These gaps coincide with the vertical
ILR and other vertical resonances. When an orbital family goes from being
stable 
to unstable, in the mentioned gaps, we get the 3-D prograde families,  i.e., 
orbits with initial conditions $(y,z,\dot y,\dot z)=(a,b,0,0)$, with $a,b \neq
0$. 
When it goes from being unstable to stable, we get the retrograde families, 
$(y,z,\dot y,\dot z)=(c,0,0,d)$, with $c,d \neq 0$, where $x$, $y$ and $z$ are 
oriented along bar's major, minor and vertical axes. 

While the vertical ILR is associated with the 
boxy/peanut bulge shapes, which develop as a result of the resonance heating of 
midplane orbits, Patsis et al. (2002b) have found that this effect 
is not limited to a particular resonance and not to the barred galaxies {\it
per se}, but operates equally well in nearly axisymmetric and/or ovally
distorted disks. Furthermore, the buckling itself is not a necessary condition
for these shapes to form, but rather accelerates the process from secular to a
dynamical timescale.

The main families of 3-D orbits which are responsible for the appearance of 
the boxy/peanut shaped bulges are BAN (prograde to the bar rotation) and ABAN
(retrograde) families found by Pfenniger \& Friedli (1991). They are called
$x1v1$ and $x1v2$, respectively, and appear as the 3-D generalizations of 
$x_1$ and $x_2$ families in the nomenclature of Skokos et al. (2002a).
In general terms,
the BAN orbits can be described as ``$\frown$" ($z > 0$) and ``$\smile$" ($z <
0$), and ABAN as ``$\infty$," when projected onto the $xz$-plane. Both families 
are $m:n:l = 2:2:1$ (i.e., two
radial oscillations for two vertical oscillations for one turn) in the
notation of Sellwood \& Wilkinson (1993).
Their projections onto the $xy$ plane have the $x_1$ orbit shapes. We trace
the planar $x_1$ and $x_2$ and the 3-D BAN/ABAN families from the bar initial
growth period. For simplicity, we have divided the simulation into a
number of characteristic time intervals and discuss them separately.

\underline{\it Before and during the first buckling, $\tau\simeq 0-2.5$~Gyr.\/} 
During this period the bar develops and buckles (Figs.~2--6). It is
geometrically thin and is not centrally concentrated, so the vertical ILR is
not present. Neither present are the planar ILRs and consequently the $x_2$
orbits. The stable 3-D orbits lie close to the midplane. As we stated above,
there is little meaning to calculate the orbits in the rapidly varying
potential without the proper time-averaging. Because the midplane of the bar
is bent in the $xz$ plane (Fig.~3), so are the $x_1$ orbits. 

\begin{figure}[ht!!!!!!]
\vbox to4.3in{\rule{0pt}{5.in}}
\includegraphics{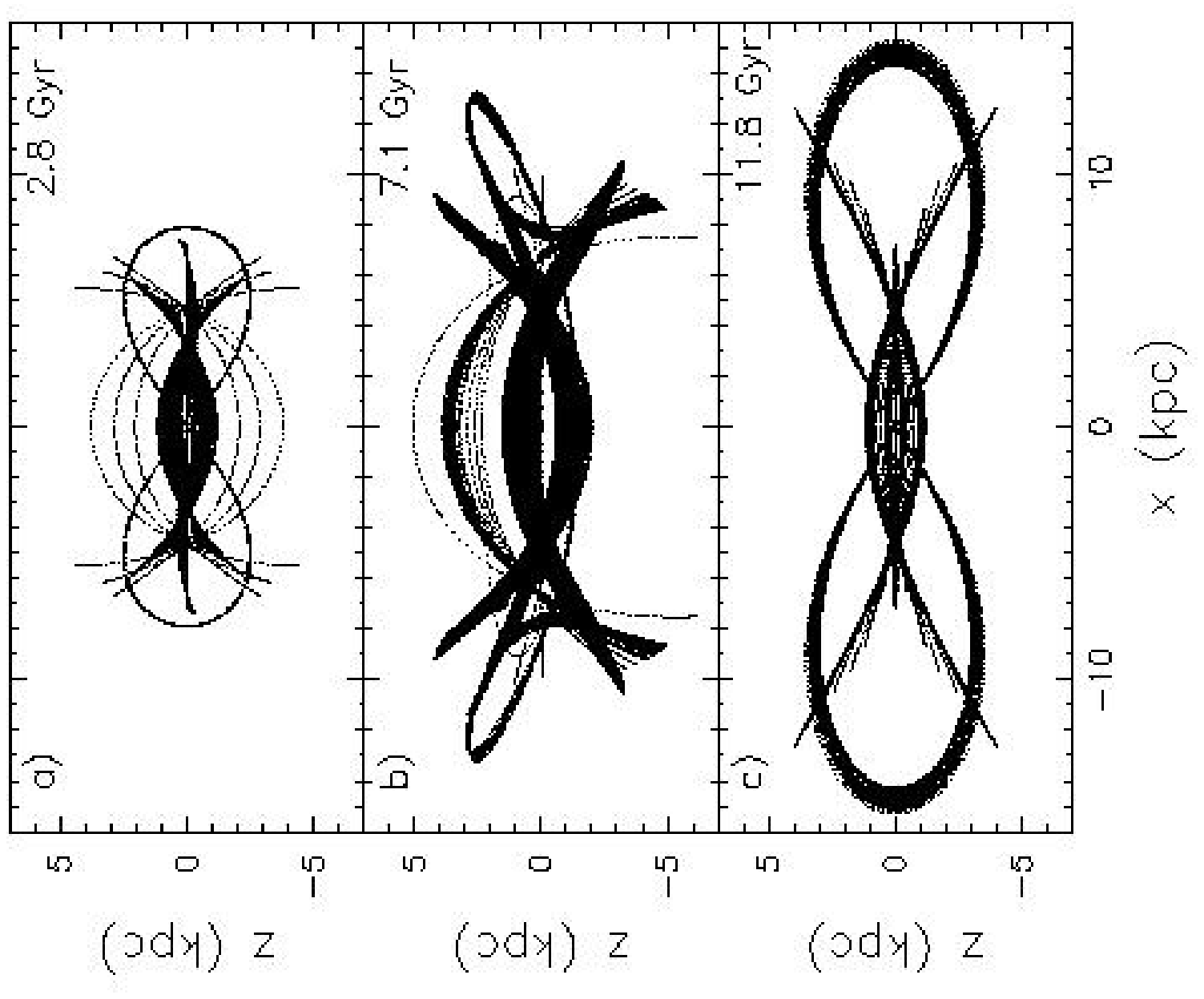}
\caption{Projection of 3-D orbits onto the $xz$ plane in three
different snapshots --- after the first buckling, during and after the
second buckling. Only {\it stable} orbits of the main families
described in the text and in Fig.~7 are plotted. The horizontal extension of
the orbits increases with time. The general shape evolves from boxy/peanut
to a vertically-asymmetric, and then to a peanut/X shape.
\label{fig:orbfig}}
\end{figure}

\underline{\it After the first buckling, $\tau\simeq 2.5-3.5$~Gyr.\/} The bar
has increased its central mass concentration and acquired its boxy/peanut shaped bulge
(see Fig.~3). As we stated above, there is little meaning to calculate the
orbits in the vertically unsymmetrized potential at the time of the first
buckling because of the rapidly varying potential. Moreover, the bar midplane
is bent in $xz$ at this time, and so are the $x_1$ orbits (Fig.~3). 
Hence, we have vertically symmetrized the potential at this time frame
(Fig.~7b, left frame). In the midplane, the $x_1$ orbits dominate and no $x_2$
family exists (Fig.~6a). Most of the $x_1$ orbits are stable, except for a
small gap, where the vertical ILR is located. At  $\tau=2.8$~Gyr the gap is at
$E_{\rm J} \sim -2.66$, where the BAN/ABAN orbits bifurcate (Fig.~7a). The
BAN/ABAN are traced (Fig.~7b,c) --- their origin in $z=0$ plane moves
gradually out with time toward higher Jacobi energy. By the end of the first
buckling it stabilizes and subsequently creeps toward higher $E_{\rm J}$ as
the bar brakes against the halo. BAN orbits appear stable from 300~pc above
the midplane, with a broad unstable gap at $z\sim 1.1-1.3$~kpc and a narrow
stability island (Fig.~7b). 
Initially, the stable part of these families is limited to a very small
height above the bar midplane, compared to the bar vertical thickness.
Gradually the stable extent of the BAN family is increasing both in $z$ and in
$x$ (Fig.~8). The ABAN family is unstable when BAN is stable and vice versa
(Fig.~7c), as first noted by Pfenniger \& Friedli (1991).  

\underline{\it Before/during the second buckling, $\tau\simeq
3.5-8$~Gyr.\/}
The bar is growing both in size and amplitude and exhibits a pronounced 
boxy profile when viewed along its minor axis. It appears symmetric with
respect to the bar midplane up to $\tau\sim 5$~Gyr (Fig.~3), although $A_{1z}$
shows that some small residual asymmetry remains between the bucklings
(Fig.~5). After $\tau\sim 5$~Gyr, the vertical symmetry 
of the bar is broken for an extended period of time $\sim 3$~Gyr 
(Figs.~3, 5). The maximal distortion happens at $\sim 7.1$~Gyr. No $x_2$ orbits 
exist during this time period and the unstable gap in $x_1$ moved to somewhat
higher energies, $\sim -2.4\lesssim E_{\rm J}\sim -1.7$, due to the bar
braking (Fig.~7a). Additional unstable gap has appeared at $-2.68\lesssim
E_{\rm J}\lesssim -2.65$, where the 3:2:1 orbit family bifurcates.  

Note that the analysis shown in the $\tau=7.1$~Gyr frame of Fig.~7 is performed
in the actual gravitational potential of the $N$-body simulation, i.e.,
without symmetrization with respect to the $z=0$ plane. Therefore, we should
not expect to find all the families existing in the symmetrized potentials.
But we should be able at least to identify the reason for the vertical
asymmetry of the particle distribution in the projection of the 3D-orbits onto
the $xz$ plane from the orbital shapes (Fig.~8b). We find the BAN family and
also the $x1v9$ family in the nomenclature of Patsis et al. (2002a).
 
The BAN family appears stable everywhere above the midplane, except around
$z\sim 2$~kpc (Fig.~7b), while ABAN is unstable everywhere and so is not shown
here. For $z\ltorder 0$, the BAN family is detached from the midplane and
 starts at $z\sim -0.5$ kpc. It forms a kind of loop in the characteristic
diagram. We do not find any stable orbit below $z\sim -2$~kpc.
The $x1v9$ family is stable only for a low range in energies and only below the
midplane, which makes it important for contributing to the vertical asymmetry
in the bar (Fig.~8b). %No ABAN orbits are found.  

\underline{\it After the second buckling, $\tau > 8$~Gyr.\/} The bar slowly
regains its vertical symmetry, and becomes symmetric after $\sim 8$~Gyr. As a
result of this instability, the bar has developed a pronounced X-shape which 
differs from the `usual' peanut shape because of the extended concave region 
between the spikes, i.e., peanuts (Fig.~8c). Orbits which appear at the 
same $z$ in the characteristic diagram but at different time frames 
differ in their projections. After the second buckling the orbits are longer
and extend higher, i.e., the convexity and concavity have significantly
changed. The vertical ILR has moved to $E_{\rm J}\sim -2.36$ by 11.8~Gyr. The
BAN orbits are stable while ABAN are unstable. Overall, the boxy/peanut bulge has
increased in size from $\sim 1-2$~kpc (after the first buckling) to $\sim
8$~kpc (after the second buckling).

The $x_2$ orbits and consequently two ILRs in the $xy$-plane appear only after 
the second buckling, $\gtorder 7.2$~Gyr. Initially occupying a small range in 
energies, their extent grows with time. The $x_1$ family is mostly stable up 
to the (mostly) unstable `shoulder' in the characteristic diagram (Fig.~7a,
right frame), which has continued to move out to higher Jacobi energies, along 
with the last stable orbit supporting the bar.  

We also use the characteristic diagrams to get an independent estimate for the
bar physical size. Unlike the case of analytical bars where this issue is
resolved
trivially, to estimate the length of the {\it live} numerical and `observed'
bars can be more difficult. For example, if the bar potential or bar
nonaxisymmetric
force is used for this purpose, one can get an erroneous result that the bar
extends beyond its corotation radius. If one relies on the density
distribution,
one can overestimate the size as well because the bar can drive a pair of
open spirals. Instead, we use the calculated orbital structure of the bar
and rely on the properties of $x_1$ family which is generally stable except
for narrow gaps (in Jacobi energy) and in the region around the corotation.
Specifically, close to the corotation, the $x_1$ curve bends upward (e.g.,
Fig.~7a; Binney \& Tremaine 1987). We use the $x$-axis extent of the last
stable
$x_1$ orbit which lies on this upward branch. It is the slow drift of unstable 
gaps and of the elbow of the $x_1$-curve towards higher $E_{\rm J}$ which leads
to the trapping of orbits and the secular increase in the bar size. A large
number
of models here and in Berentzen et al. (2005) have been used for the comparison
with the isodensities fit to the bar (Section~3). 

The bar amplitude $A_2$  shows that the outer half of the bar practically
dissolves 
in the first buckling and weakens substantially in the recurrent buckling
(Fig.~2). 
The ellipse fit to the bar reflects this by a dramatic decrease in the bar
length 
from $\sim 11$~kpc to $\sim 6$~kpc in radius. Unfortunately, it is more tricky
to 
estimate the bar size from the orbital analysis at this time --- the underlying
potential is time-dependent while the analysis is performed in the
frozen potential. Nevertheless, both methods agree before and after the 
the first buckling and capture the change in the growth rate of the bar length
after the second buckling. To summarize, the bar size decreases during
the first buckling and levels-off during the second one. 

The simple explanation for the success of the orbital analysis method is that
it is the most self-consistent method known to
us and relies on the orbit trapping by the stable periodic family whose
existence basically defines the bar. We note that the middle frame
of Fig.~7a does show a large unstable gap just below the elbow of $x_1$ curve.
However, we note once more that the potential in this frame has not been
vertically
symmetrized and that the weakening of the bar during this time of buckling
is in fact a direct consequence of this gap.

In summary, the characteristic diagrams and their respective orbital
families reflect both the dynamical and secular evolution of the bar shown
in both Fig.~3 and the Animation Sequence. The ustable gaps, where the vertical
families (BAN/ABAN) bifurcate, move toward higher Jacobi energies while
the orbits become more extended along the bar axes.
The orbital families evolve not just quantitatively but also qualitatively
--- this shows up in the changing shape of the bar. Although we do not measure
the population of the different families, the bar shapes appear to be governed
mainly by the BAN family, which evolves by changing the concavity of its
orbits, thus giving the
bar the pronounced peanut/boxy/X shapes when viewed along the minor axis. The
change in the orbital structure of the bar supports our observation, based on
independent arguments, that there is a recurrent buckling.
At the secondary buckling time, the vertical families differ above and below
the plane. The former are stable orbits extending more along the $x$-axis and
also giving it more of a boxy shape. The latter ones are stable orbits which
give the bar more of a peanut shape. The appearance of the $x_2$ family
confirms that the ILRs form only after the second buckling of the bar.  

%%%%%%%%%%%%%%%%%%%%%%%%%%%%%%%%%%%%%%%%%%%%%%%%%%%%%%%%%%%%%%%%%%%
\section{Disk--Halo Angular Momentum Exchange: the Role of the Resonances}
%%%%%%%%%%%%%%%%%%%%%%%%%%%%%%%%%%%%%%%%%%%%%%%%%%%%%%%%%%%%%%%%%%%

The disk region which gives rise to a bar is known to lose its angular momentum
($J$). In principle, a number of components in a galaxy can acquire this
momentum, namely, the outer disk, the bulge, and the halo. While the
former two components are able to store $J$, the responsive (i.e., live) halo
can serve as a particularly large angular momentum sink due to its large mass 
and low $J$.
Athanassoula (2002b, 2003) has shown that the disk-halo interaction
is mediated by lower resonances. In order to understand more fully the complex 
behavior exhibited by the model, such as the initial bar growth, its recurrent 
bucklings, and subsequent secular growth --- we find it instructive to
determine the balance of the angular momentum and its evolution. We pay 
particular attention to the resonant interactions between the model
components. 

\begin{figure}[ht!!!!!!]
\vbox to5.0in{\rule{0pt}{3.3in}}
\includegraphics{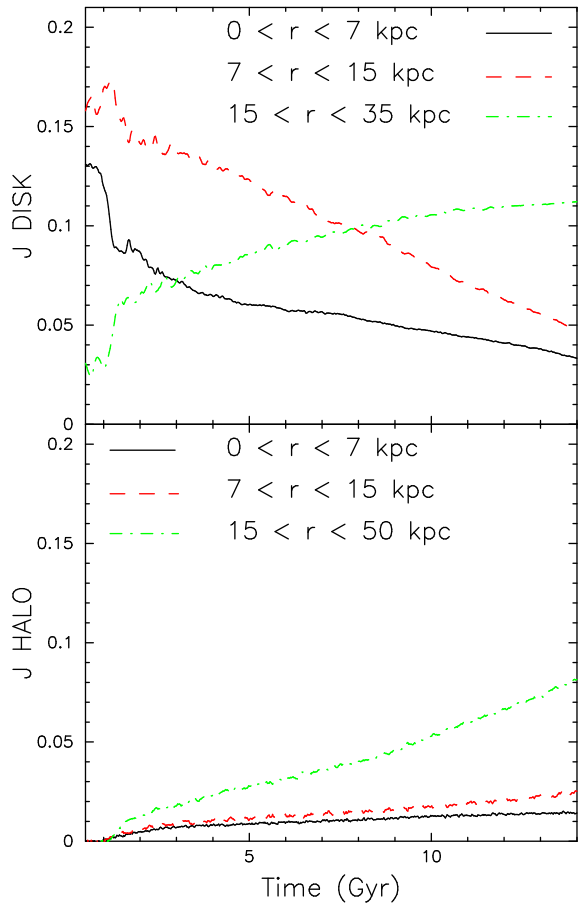}
\caption{Evolution of angular momentum in the various regions of
the disk (upper) and the halo (lower). The specified radii are given in
cylindrical geometry. 
\label{fig:a2ampl}
}
\end{figure}

\begin{figure*}[ht!!!!!!]
\vbox to5.0in{\rule{0pt}{3.3in}}
\includegraphics{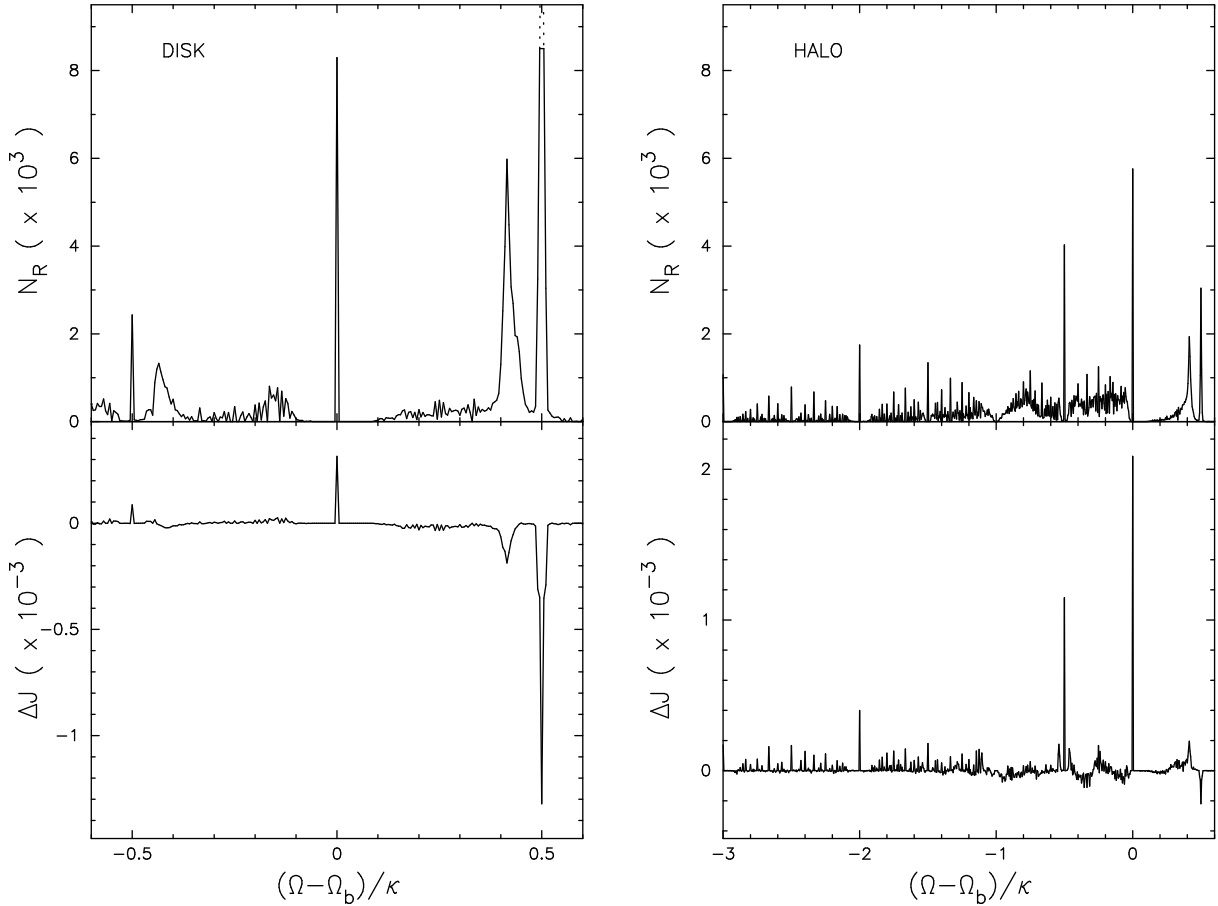}
\caption{Resonance interaction in the live disk-halo system. {\it
Upper:} histogram of particle distribution $N_{\rm R}$ with the principal
frequency ratio $\eta\equiv (\Omega-\Omega_b)/\kappa$ for the disk and
(rotating) boxy/peanut bulge 
(left) and halo (right), about $1.5\times 10^5$ randomly picked particles each,
at $\tau=2.8$~Gyr. The cusp at the ILR ($\eta=0.5$) for the disk is shown only
partially --- it extends to $\sim 39\times 10^3$. The broad peaks near
$\eta\sim \pm 0.4$ are made mostly of near-resonant bulge particles injected
during the first buckling. 
{\it Lower:} angular momentum difference, $\Delta J$, between the particles at
$\tau=5.2$~Gyr and $\tau=2.8$~Gyr in the disk (left) and halo (right) as a
function of $\eta$ at time 5.2~Gyr. The lower resonances, such as ILR ($\eta =
0.5$), the
corotation ($\eta=0$), OLR ($\eta=-0.5$), 1:3 ($\eta=\pm 0.33$), 1:4
($\eta=0.25$), etc. are clearly visible. The ILR dominates the loss of the
angular momentum in the disk, and the corotation dominates the absorption of
the momentum by the halo (more in the text). The frequency ratio is binned
into intervals of $\Delta\eta=0.005$. Note that $\eta$ scales are different for
the disk and the halo.
\label{fig:a2ampl}
}
\end{figure*}

We first determine the overall $J$ balance in the different regions
of the disk and the halo. Fig.~9 shows the total angular momentum evolution for
three different regions in the disk and halo: inner ($0-7$~kpc),
intermediate ($7-15$~kpc) and outer ($15-35$~kpc or 15-50~kpc) regions. The
inner and intermediate regions are those hosting the bar for various times
during its evolution. The time of the bar formation is characterized by a
substantial loss of $J$ in the inner disk. This angular momentum is
redistributed to the intermediate and especially to the outer disk (Fig.~8). 
During the first buckling, there is some indication that the inner
disk gains some angular momentum, and we return to this issue in section~6.2.
The subsequent growth of the bar is accompanied by a slow $J$ growth in the
outer disk and the halo. Interestingly, after the second
buckling $J$ stops growing in the outer disk, while the outer halo picks-up $J$ 
at an increasing rate. There is a general loss of
angular momentum from the disk and there is a general gain in the halo. 
The loss of $J$ in the disk correlates with the slowdown
of the bar, which is known to anti-correlate with the growth of the bar
(e.g., Athanassoula 2003).  

Next, we calculate the contribution of the resonances between disk and halo 
particles with the bar's $\Omega_{\rm b}$ to the angular momentum
exchange between the disk and the halo. We generally follow the procedure
described in Athanassoula (2002b). The principal frequences, $\Omega$
and $\kappa$ of azimuthal motion and of radial oscillations of the 3-D
orbits, have been determined by means of spectral analysis (Binney
\& Spergel 1982). We freeze the gravitational potential in the model at 
times $\tau=2.8$~Gyr and 5.2~Gyr, but allow the bar to tumble with its original
$\Omega_{\rm b}$, and integrate the orbits of $1.5\times 10^5$ randomly-picked
particles in the disk and the same amount in the halo for an additional
9.4~Gyr, about 20 bar tumblings.
Results are plotted in Fig.~10 as a function of the frequency ratio $\eta\equiv 
(\Omega-\Omega_{\rm b})/\kappa$, where $N_{\rm R}$ particles are binned in the
intervals of $\Delta\eta=0.005$ (top frames). The difference $\Delta J$ in 
angular momentum between the two times is given as a function of $\eta$ at 
$\tau=5.2$~Gyr (bottom frames). The
major resonances are located at $\eta=\pm 0.5$ (ILR and OLR), $\pm 1:3, 1:4,$ 
etc., with the positive $\eta$ indicating the resonances inside the corotation 
radius and $\eta<0$ --- outside the corotation. The corotation resonance 
corresponds to $\eta=0$. 

The resonant and near-resonant particle distributions $N_{\rm R}$ at the time
$\tau=2.8$~Gyr are clearly non-uniform and are permeated by numerous
resonances. 
The trapping of orbits by the resonances is also evident. The dominant
resonance in the disk appears to be
the ILR which traps the largest number of particles and therefore facilitates 
the loss of the angular momentum from the disk (lower panels of Fig.~10). Note,
that we have cut $N_{\rm R}$ for this resonance to keep the same scales for
disk and halo diagrams --- it extends to $39\times 10^3$. The $x_1$ within the
CR and later on $x_2$ (if populated) orbits, as well as BAN/ABAN orbits are 
trapped within this 
resonance. The broad asymmetric peak (Fig.~10a, left panel) in
the range of $\eta\sim \pm 0.35-045$ is made out of near-resonant orbits in
the (rotating) boxy/peanut bulge, many of them near resonant BAN orbits. Those have been 
injected during the first buckling. They appear on both sides of the CR because 
$J$ along these orbits oscillates wildly, $\Delta J\sim J$, and it is sometimes
difficult to disentangle the prograde from retrograde ones among them.
For comparison, we repeated this procedure at $\tau=11.8$~Gyr and confirmed
that more than half of the particles in the above broad peaks have been
trapped by the nearby ILR and OLR by that time. The resonances for $\eta 
\ltorder 0$, with CR and OLR as the next strongest ones, absorb the angular 
momentum and their $\Delta J$ are all positive. 

The halo is dominated by the corotation resonance which absorbs $J$ from the
disk. Overall, the disk resonances emit and the halo resonances absorb $J$.
This behavior was demonstrated by Athanassoula (2002b, 2003), where similar 
plots to Fig.~10 were given and there is a very good agreement between 
our results. The only difference we 
find is that the halo ILR is actually losing $J$ in our model, albeit a small 
amount, unlike in Athanassoula's models. The modeled halo extends
well beyond the disk and its
outer part appears to be actively storing $J$, up to $\eta\sim -3$. 
This effect is confirmed by the overall $J$ evolution in the halo (Fig.~9).
We find that the trapping of orbits by the resonances is robust ---
almost all the particles trapped at $\tau=2.8$~Gyr remain trapped at 5.2~Gyr. 
The ability
of the barred disk to transfer its angular momentum to the halo while $J$
saturates in the outer disk, explains why the bar is able to grow during
this time interval as exhibited by the $A_2$ amplitude and the size of the bar
(e.g., Figs.~2--5). It is this growth of the bar which is ultimately
responsible for the recurrent buckling phenomenon analyzed here.
 
%%%%%%%%%%%%%%%%%%%%%%%%%%%%%%%%%%%%%%%%%%%%%%%%%%%%%%%%%%%%%%%%%
\section{Discussion}
%%%%%%%%%%%%%%%%%%%%%%%%%%%%%%%%%%%%%%%%%%%%%%%%%%%%%%%%%%%%%%%%%

 We have modeled the evolution of collisionless (stellar) bars embedded in
responsive (live) axisymmetric dark matter halos using high-resolution
$N$-body simulations. We find that (1) bars experience secular growth
over the simulation ($\sim$~Hubble) time, except (2) during well
defined time periods when the bars encounter spontaneous breaks of
vertical symmetry, so-called buckling (or firehose) instability (e.g.,
Fig.~3 and the Animation Sequence~1). 
We detect such a recurrent buckling during which the bar, and especially its
outer half, weakens substantially, but the growth is resumed subsequently.
Two different techniques have been used to measure the bar size, namely,
the nonlinear orbital analysis and the isodensity ellipse fitting. The bar 
strength has been measured using the amplitude of $m=2$ mode, and the
bar vertical asymmetry --- using the vertical $m=1$ mode. Moreover,
we have analyzed the bar evolution and its buckling periods by means of 
the nonlinear orbit analysis and the ratio of vertical-to-planar velocity 
dispersions. Finally, we have examined the bar development in terms of the 
angular momentum redistribution between various components in the disk-halo
system mediated by the resonant interactions using orbital spectral
analysis.

Low resolution $N$-body simulations exhibit fast (e.g., during one disk
rotation) stellar bar growth, followed by a vertical buckling and
secular weakening. High-resolution simulations with $N\gtorder 10^{5-6}$
show a more complex evolution. They allow for modeling the disks embedded
in live halos and account for the resonant interactions between
the halo and the disk, adding a new and crucial element to the simulations in 
the form of 
angular momentum transfer between these components. The main difference with
the previous low-resolution models and that presented in this work is the 
ability of the stellar bar
to strengthen again after its original weakening following the buckling --- a
process which leads to the bar growth and consequently to its recurrent 
buckling.

\subsection{Disk-Halo Resonant Interactions}

The bar growth has been associated with the existence of `sinks'
of angular momentum located elsewhere. Athanassoula (2002b, 2003) has shown
that a
galactic halo can play such a role and absorb large quantities of angular
momentum from the disk/bar region. This effect appears to completely invert
the original suggestion by Ostriker \& Peebles (1973) about the stabilizing
function of dark matter halos in disk galaxies against the bar formation
instability. Angular momentum redistribution, at least in principle, can have 
contributions both from the
non-resonant and resonant interactions between the bar and orbits in the halo.
To capture the latter requires large $N$ to stabilize
the population of resonant particles (e.g., Weinberg \& Katz 2002).

It is therefore important to demonstrate that
resonant particles are indeed responsible for the angular momentum transfer
in the first place. We start with the relative contributions to the $J$
transfer 
between the disk and
the halo and between the inner (i.e., bar unstable $\sim 15$~kpc) and outer
disks (Fig.~9). Although the efficiency of this redistribution is of course
model dependent, we nevertheless quote the numbers, assuming that to a certain 
degree, they are representative. We find that $\sim 39\%$ of the angular
momentum in the disk is lost to the halo during the evolution. The bar
unstable region, $\ltorder 15$~kpc has lost about 71\% of its original $J$, of
which about 38\% went to the outer disk and the rest was
absorbed by the halo.
The most intensive flow of $J$ to the outer disk happens in the early stage
which ends with the first buckling. The halo particles have much larger
dispersion velocities than the disk and until the bar fully develops, their
resonant interaction with the $m=2$ asymmetry in the disk is virtually
non-existent. 

After the first buckling, it is the halo which absorbs most of the angular
momentum lost by the bar region and from Fig.~9 (lower frames) it is clear
that the lion share of this exchange is resonant and mediated by the CR and
the OLR in the halo and by the ILR and the CR in the disk, with some
contribution from the numerous minor resonances. This results in the increase
of the
bar size and strength, and leads to the gradual decrease in the ratio of the
velocity dispersion, $\sigma_{\rm z}^2/\sigma_{\rm r}^2$, in a close analogy
with the disk evolution preceeding the first buckling. The secondary buckling
of the bar, therefore, can be directly traced to its robust growth in our
model.  
 
The issue of a bar dissolution vs. growth has become controversial recently.
While it is beyond the scope of this work, indirectly we do touch it.
First, we confirm the results of Martinez-Valpuesta \& Shlosman (2004) that
vertical bucklings do not destroy the bar, unlike suggested by Raha et al.
(1991; see also Sellwood \& Wilkinson 1993). Next, in various axisymmetric
models, here and in Berentzen et al. (2005), a robust growth in the
bar is observed after the first buckling, supplemented by the angular 
momentum exchange
between the disk and the halo. Athanassoula (2003) has analyzed the effect of
initial conditions on this $J$ redistribution, namely, of the initial
disk-to-halo mass ratio and of the Toomre's Q-parameter --- the angular
momentum exchange varied from $\sim 2\%$ to $\sim 35\%$ with a clear
correlation between the amount of the exchanged momentum and the size of the
bar. In comparison, our models have a halo-to-disk mass
ratio of unity within the central 10~kpc, $Q\sim 1.5$, and a halo core of $\sim
2$~kpc. They lie between the models MH1, M$\gamma$3-4 and MQ4 of Athanassoula.

Interestingly, Valenzuela \& Klypin (2003) argued that it is the mass and force
resolutions of numerical models which dictate the efficiency
of $J$ transfer to the halo and hence play an important role in the evolution
of the bar. Insufficient resolution results in an excessive growth of
numerical bars and an excessive decrease in the bar pattern speed. However, it
seems rather that the physical conditions mentioned above take priority ---
their high resolution models have extended and hot halos. High dispersion
velocities in the halo affect the particle trapping by the resonances and so
are expected to lower substantially the $J$ transfer.  

\subsection{Bar size evolution}

If one neglects the dissipative component in a modeled galaxy, the bar pattern
speed
decreases secularly, except during brief time intervals
of internal instabilities in the bar itself. At
least in numerical models of stellar bars, their overall slowdown is
accompanied by an increase in the bar length, so the bar roughly extends to
its corotation radius which increases with time (Athanassoula 1992). 
Alternative theoretical models of bars terminating at the ILR exist
(Lynden-Bell 
1979), but difficulties remain in actually reproducing them in numerical 
simulations (but see e.g., Polyachenko \& Polyacnenko 1994 for the so-called
slow
bars). A sole exception consists of a system of 
nested bars, where the inner (nuclear) bars form within the ILR (e.g.,
Shlosman, Frank \& Begelman 1989;
Friedli \& Martinet 1993; Englmaier \& Shlosman 2004). However, a dissipative
component is required to be present for self-consistency in this case.
How exactly the bar traps the disk orbits in order to increase its length
is not known at present. 

We have demosntrated that while the size of the bar, $r_{\rm bar}$, drops during 
the first buckling (Fig.~2a), the ratio of $r_{\rm CR}/r_{\rm bar}$ (Fig.~1b) 
does {\it not} increase 
dramatically above the range of $1.2\pm 0.2$ determined by Athanassoula (1992)
to fit the observed shapes of the offset dust lanes in barred galaxies. 
Amazingly, the reason for this is that $r_{\rm CR}$ drops as well due to the
sudden increase in the bar pattern speed in the same time interval. 
This is not unexpected due to a clear trend between $\Omega_{\rm b}$
and $A_2$ (Athanassoula 2003) and a long known fact that the bar lengthens
with its slowdown in numerical simulations. What is new here is that we find 
that $\Omega_{\rm b}$ increases sharply during the first buckling and this increase
apparently correlates with the sudden decrease in the bar length. Hence
the relation between between $\Omega_{\rm b}$, $A_2$ and $r_{\rm bar}$ 
appears to be more fundamental than anticipated and holds for decelerating and
accelerating bars altogether.     

We associate the continued bar growth with the ability of the halo to absorb
angular momentum from the disk region lying within the corotation radius ---
a region which itself expands with time. This process leads to a stronger bar
and can be seen as a counterbalance to a number of other processes which have
been discussed in the literature within the framework of bar dissolution. 
Recent observational results from Galaxy Evolution from Morphology and SEDs
(GEMS, Rix et al. 2004) survey have shown that the bar size and axial ratio
distributions at intermediate redshifts of $z\sim 0.2-1$ are compatible with 
those in the local Universe (Jogee et al. 2004; Elmegreen et al. 2004; 
Sheth et al. 2003). Our models which exhibit a slow secular bar growth over the
Hubble time are in general agreement with these results. They do lack the 
gaseous component which can dramatically shorten the bar life cycle (e.g.,
Bournaud \& Combes 2002). However, it is difficult to understand how the
short-lived (e.g., less than 1.5~Gyr [Bournaud \& Combes 2002; Combes 2005])
bars can form at the same rate and with the same sizes and axial ratios
they are being destroyed by the gas. Yet, taken at the face value, in order 
to agree
with the GEMS results, the bars must preserve their size and strength
distributions nearly unchanged over the last 8~Gyr. Clearly, the precise role
of gas in the bar evolution must still be determined.  

%%%%%%%%%%%%%%%%%%%%%%%%%%%%%%%%%%%%%%%%%%%%%%%%%%%%%%%%%%%%%%%%%%%%%
\subsection{Bar shape evolution}

Next we focus on some aspects of the bar 3-D shape evolution in the model, 
specifically on the axial ratio in the bar midplane and on the bar symmetry
in the vertical plane. The midplane axial ratio, or alternatively the
bar ellipticity $\epsilon$, is measured from the ellipse fitting (Fig.~4c)
and agrees with the evolution of the bar amplitude $A_2$. At the same time,
the variations in $\epsilon$ appear less dramatic than in $A_2$, although
recurrent bucklings are clearly visible as a decrease or saturation.   

The ratio $r_{\rm CR}/r_{\rm bar}$ has also dynamical implications
for the bar. For example, it determines the shape of the offset dust
lanes in barred galaxies, which in turn delineate shocks in the gas flow
(Athanassoula 1992). The strength of the underlying shocks determines the 
gas inflow towards the central kpc. A bar which is considerably weakened 
will slow down the radial gas inflow.  The observed shapes constrain the ratio 
$r_{\rm CR}/r_{\rm bar}$ to $1.2\pm 0.2$. The modeled ratio (Fig.~4b) 
typically falls within the required limits except during the first buckling 
when it is higher, $\sim 1.5$.   

%FIG.10 ANSAE
\begin{figure}[ht!!!!!!]
\vbox to3.5in{\rule{0pt}{3.in}}
\includegraphics{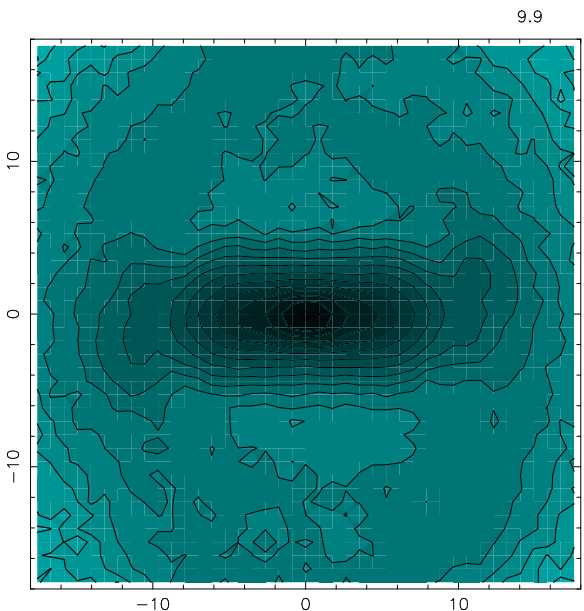}
\caption{ {\it Ansae} at time $\tau=9.9$~Gyr. A face-on view of the
modeled barred disk. The gray-scale contours represent the surface density and
are spaced logarithmically. The box size is 18~kpc $\times$ 18~kpc.
\label{fig:a2ampl}
}
\end{figure}

After the second buckling, the bar shows the so-called {\it ansae} (handles) 
on both its ends. We observe them as characteristic density
enhancements in the face-on (Fig.~11) or edge-on (Fig.~3) disks. Athanassoula
(2001) related the appearance of {\it ansae} to initial conditions in the
models (e.g., the halo-to-disk mass ratio). It remains unknown why they 
appear at this particular evolutionary stage, $\tau > 8$~Gyr. The {\it ansae} 
can be seen in some early-type barred galaxies, e.g., NGC~4262, NGC~2859 and 
NGC~2950 (Sandage 1961), NGC~4151 (Mundell \& Shone 1999), ESO 509-98 (Buta et
al. 1998), etc.

Vertically, the bar evolves from a geometrically thin configuration,
similar to the disk hosting it. The vertical bar buckling, when viewed along
the bar's minor axis, shows a rapidly evolving bending which relaxes to a boxy
shaped bulge, i.e., bulge with flat or mildly convex isodensities (Fig.~8a). 
Subsequently, the bulge acquires a peanut shape. During the second buckling, 
for $\sim 3$~Gyr,
the vertical asymmetry persists with one-sided boxy and peanut symmetries 
which derive from the asymmetric BAN family (Fig.~8b). With the asymmetry 
washed out, the boxy/peanut bulge/bar has a pronounced 
X-shape: two pointed spikes with a large concave region in between (Fig.~8c). 
The X-shapes have been seen 
before in numerical simulations (e.g., Athanassoula \& Misiriotis 2002) and
have been observed as well (e.g., NGC~4845, NGC~1381, IC~4767 (Whitmore
\& Bell 1988), IC~3370 (Jarvis 1987) and AM~1025-401 (Arp \& Madore
1987; Bureau \& Freeman 1999; Patsis et al. 2002a). Mihos et al. (1995)
proposed a merging scenario for the formation of the X-shaped bulges, 
e.g., a minor merger for Hickson~87a galaxy. This merger triggers the bar
instability in the disk, followed by the buckling and the X-shaped bulge, 
when viewed from a specific
aspect angle. In principle, there may be more than one cause for the 
bar buckling and for the formation of peanut/boxy/X-shaped bulges. But
when taken together, a high observed frequency of these bulges (L\"utticke et
al. 2000) and the need for only mild asymmetry in the disk for their
appearance (Patsis et al. 2002b) may hint to their intrinsic origin.  

%%%%%%%%%%%%%%%%%%%%%%%%%%%%%%%%%%%%%%%%%%%%%%%%%%%%%%%%%%%%%%%%%%%%%%%%%%%%%
\subsection{Vertical asymmetry of the bar}

The vertical asymmetry of the modeled bar has been detected during its 
recurrent bucklings, for about 1~Gyr and $\sim 3$~Gyr, respectively 
(Figs.~3, 5). Its characteristic shape, in principle, is detectable by 
observations of edge-on galaxies (e.g., Fig.~3 
and the Animation Seq.~1), especially during the second buckling due
to its prolonged period. Observationally, the bucklings
differ when viewed edge-on along the bar minor axis.
The main difference is the location of the maximal asymmetry --- it is
close to the rotation axis during the first buckling and around the
middle region of the bar during the second buckling. 

L\"utticke et al. (2000) present first statistics of edge-on galaxies 
with the boxy/peanut bulges and show some isophote fits (their Fig.~2). 
They distinguish between peanut-shaped and boxy-shaped bulges.
A total of 27\% of the sample of 734 galaxies with boxy/peanut bulges 
have bulges which are either close to boxy or, due to low resolution,
could not be distinguished between boxy and peanuts. Some of these
bulges show a mild vertical asymmetry similar to that found during
the second buckling, e.g., NGC~4289 ---  a good example of a possible 
observed secondary buckling. Of course, to corroborate the numerical 
simulations one needs a large, statistically 
significant sample with high resolution. The main difficulty lies in
the unambigious determination of the inclination angle of a galactic disk --- 
it must be edge-on within $\pm 5^\circ-7^\circ$.

From a theoretical point of view, it is unclear how wide-spread is the
recurrent buckling in stellar bars --- how sensitive it is to disk
and especially halo parameters, such as mass distribution and dispersion
velocities. While the condition for a second buckling seem to
be directly related to the ability of the bar to strengthen after
the initial weakening, this process can depend on a number of additional 
parameters. For example, Athanassoula (2003) has found that the angular
momentum transfer between the disk and the halo depends on the mass
distribution in the halo and weakens substantially for hotter halos.
Berentzen et al. (2005) have detected a recurrent buckling in
their LS2 model with 2~kpc flat core (live) halo with a logarithmic potential.
At the time of the second buckling, the bar
in this model appears somewhat stronger than in the first buckling. 
Moreover, the secondary buckling can be apparently seen in Fig.~4 of 
O'Neil \& Dubinski (2003) and also in Fig.~11 of Valenzuela \&
Klypin (2003), the latter based on the evolution of the bar pattern speed,
strength and
the angular momentum rate change. Secondary buckling is also
present in some models of L. Athanassoula (private communication
2005). In all the cases they went
unnoticed. Lastly, we comment on the addition of the
dissipative component to the stellar disk (Berentzen et al. 1998). 
The effect of the clumpy gas component is to weaken the buckling instability,
but without quantifying the degree of clumpiness within the central kpc,
it is not clear whether the isothermal equation of state used has led in fact
to over-damping of this instability. 

\section{Conclusions}

To summarize this work, we have studied the long-term stellar bar evolution
in a high-resolution self-consistent model of a disk and a responsive 
halo. We find that a developing bar goes through the vertical buckling 
instability which weakens it and dissolves its outer half. Subsequently, the 
bar experiences a renewed growth which leads to a recurrent buckling. This 
evolution is driven by the resonant interaction between the barred disk and the 
surrounding halo --- we quantify this effect by means of the spectral analysis 
of individual orbits in the disk and the halo and show that the halo particles
are trapped by numerous lower resonances with the bar and that this
trapping is robust. During these periods of recurrent instability, and 
especially during the slower second buckling, the bar remains 
vertically asymmetric for a prolonged $\sim 3$~Gyr time interval --- which in
principle can be detected observationally.
However, two issues can potentially complicate this detection. First, it is not
clear how widespread are the conditions favorable for the recurrent bar
growth, although we have detected it in a number of models with different
initial conditions. Second, while it was shown that a clumpy gaseous component
with an isothermal equation of state in the disk will weaken this instability,
the effect of a realistic ISM was never estimated. A statistically significant
sample of (nearly) edge-on galaxies is required to test the prediction of a
prolonged vertical asymmetry.
 
We also find that the secular bar growth and the triggered buckling
instabilities lead to pronounced changes in the bulge shape --- it grows both
radially and vertically, acquiring a peanut, a boxy and finally the X-shaped
appearance. While the bar size approximately follows the 4:1 
(Ultra-Harmonic) resonance in the disk, the boxy/peanut bulge size appears to be guided 
by the vertical ILR. Concurrently, the bar is going through a structural
evolution --- new families of 3-D periodic orbits appear (or become more
pronounced) after the bucklings.  
 
Finally, we find that the bar strength correlates with its pattern speed, in
both
strengthening and weakening bars. While it was known already that a stellar bar
becomes stronger as it slows down, we detect the reverse trend as well ---
bars that weaken during the buckling speed up their tumbling. Moreover, 
the bar size appears to be sensitive to these changes --- slowing down bars
become longer, while speeding up bars shorten.   

\acknowledgments
We gratefully acknowledge insightfull discussions with Lia Athanassoula, Ingo
Berentzen, Amr El-Zant, Johan Knapen and Alar Toomre. This work has been 
partially supported by grants
from NASA/LTSA5-13063, NASA/ATP NAG 5-10823, HST AR-09546 and 10284 (I.S.), 
and by NSF AST-0206251 (C.H. \& I.S.). I.M. acknowledges support from 
PPARC. Simulations and orbital analysis have been performed on a dedicated 
Linux Cluster and we thank Brian Doyle for technical support.

\end{document}